\documentclass[aps,prl,twocolumn,reprint]{revtex4-1}
\usepackage{amsmath,amsthm,amsfonts,ulem}
\usepackage{graphicx, color}
\begin{document}
\title{Undoing a quantum measurement}

\author{Philipp Schindler$^1$} 
\author{Thomas Monz$^1$}
\email{thomas.monz@uibk.ac.at} 
\author{Daniel Nigg$^1$} 
\author{Julio  T. Barreiro$^{1}$} \altaffiliation{Current address: Fakult\"at f\"ur
  Physik, Ludwig-Maximilians-Universit\"at \& Max-Planck Institute of
  Quantum Optics}  
\author{Esteban A. Martinez$^1$} 
\author{Matthias  F. Brandl$^1$} 
\author{Michael Chwalla$^{1,2}$} 
\author{Markus  Hennrich$^{1}$} 
\author{Rainer Blatt$^{1,2}$}

\affiliation{$^1$Institut f\"ur Experimentalphysik, Universit\"at
Innsbruck, Technikerstrasse 25, A--6020 Innsbruck,
Austria\\$^2$Institut f\"ur Quantenoptik und Quanteninformation
der \"Osterreichischen Akademie der Wissenschaften,
Technikerstrasse 21a, A--6020 Innsbruck, Austria}

\begin{abstract}
  In general, a quantum measurement yields an undetermined answer and
  alters the system to be consistent with the measurement result. This
  process maps multiple initial states into a single state and thus
  cannot be reversed.  This has important implications in quantum
  information processing, where errors can be interpreted as
  measurements.  Therefore, it seems that it is impossible to correct
  errors in a quantum information processor, but protocols exist that
  are capable of eliminating them if they affect only part of the
  system. In this work we present the deterministic reversal of a
  fully projective measurement on a single particle, enabled by a
  quantum error-correction protocol that distributes the information
  over three particles.
  % Quantum mechanics teaches us that a quantum system may be in a
  % superposition of multiple states at once. While this was, and still
  % is, subject to discussions, measurements in quantum mechanics become
  % even more counter-intuitive: A measurement that is able to
  % distinguish between the constituents of the superposition gives only
  % a single result and alters the state such that it is consistent with
  % the measurement outcome. The identical state after the measurement
  % can originate from multiple different initial states and thus the
  % measurement cannot be undone which has important implications in
  % quantum cryptography and quantum information processing.
  % Interestingly, any error that occurs in a quantum information
  % processor can also be interpreted as a measurement and thus it seems
  % that it is impossible to correct occurring errors. Nevertheless,
  % quantum error correction protocols exist that encode the information
  % of a single particle into an entangled state of multiple
  % particles. In this work we present the deterministic reversal of a
  % fully projective measurement on a single particle, utilizing a
  % quantum error-correction protocol that distributes the information
  % over three particles. This work emphasizes that a measurement on a
  % single particle actually can be undone, which seemingly contradicts
  % the foundations of quantum mechanics. This contradiction can be
  % easily solved if one takes into account that no information about
  % the encoded quantum state is gained when measuring a single particle
  % in a protected register.
\end{abstract}
\maketitle

Measurements on a quantum system irreversibly project the system onto
a measurement eigenstate regardless of the state of the
system. Copying an unknown quantum state is thus impossible because
learning about a state without destroying it is prohibited by the
no-cloning theorem\cite{Wootters1982Single}.  At first, this seems to
be a roadblock for correcting errors in quantum information
processors. However, the quantum information can be encoded redundantly in
multiple particles and subsequently used by quantum error correction
(QEC)
techniques~\cite{steane96,shor_good,ions04,beyond09,photons,nmr3}.
% Measurements on a quantum system project the system into an eigenstate
% of the measurement operator which is an irreversible process for any
% state that is not an eigenstate of the measurement operator.
% It follows that it is impossible to copy an unknown state, since there
% is no way to learn about a state without destroying it as formulated
% in the no-cloning theorem\cite{Wootters1982Single}. 
% This seems to be a
% roadblock for correcting errors in quantum information
% processors. However, it is possible to store information redundantly
% in multiple particles and use this to correct errors, as described and
% actually demonstrated in the framework of quantum error
% correction~\cite{steane96,shor_good,ions04,beyond09,photons,nmr3}.
When one interprets errors as measurements, it becomes clear that
such protocols are able to reverse a partial measurement on the
system. In experimental realizations of error correction procedures,
the effect of the measurement is implicitly reversed but its outcome
remains unknown. Previous realizations of measurement reversal with
known outcomes have been performed in the context of weak measurements
where the measurement and its reversal are probabilistic
processes\cite{Kim2009Reversing,Kim2011Protecting,Katz2008Reversal,Katz2006Coherent}.
We will show that it is possible to deterministically reverse
measurements on a single particle.

We consider a system of three two-level atoms where each can be
described as a qubit with the basis states $|0\rangle,|1\rangle$. An
arbitrary pure single-qubit quantum state is given by $|\psi \rangle =
\alpha |0\rangle + \beta |1\rangle$ with $|\alpha|^2 + |\beta|^2=1$
and $\alpha, \beta \in \mathbb{C}$. In the used error-correction
protocol, the information of a single (system) qubit is distributed
over three qubits by storing the information redundantly in the state
$\alpha |000\rangle + \beta |111\rangle$.  This encoding is able to
correct a single bit-flip by performing a majority vote and is known
as the repetition code~\cite{nielsen_chuang}.

\begin{figure}[h]
\centering
\includegraphics[width=8.9cm]{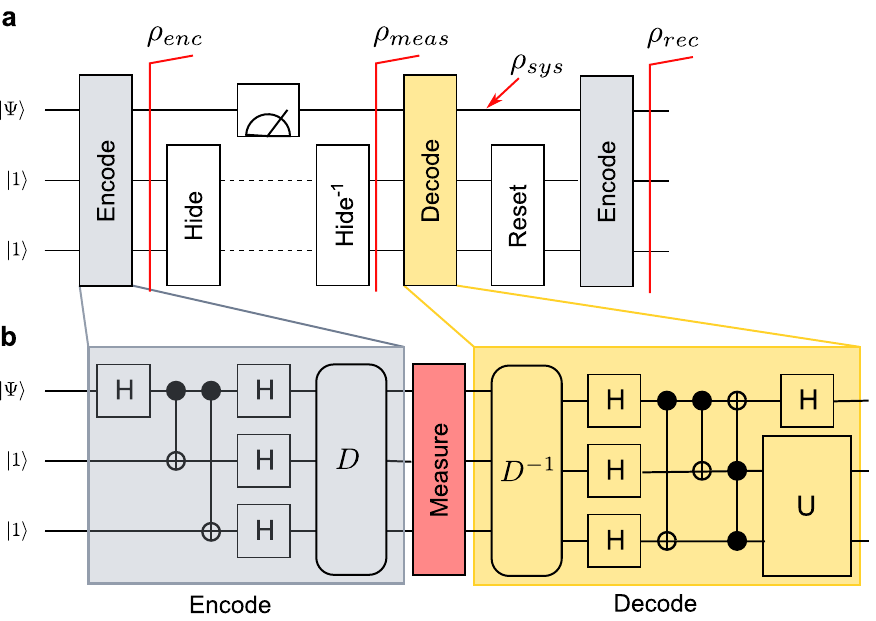}
\caption{(a) Schematic circuit of undoing a quantum measurement.
  $\rho_{enc}$ is the encoded state of the register, $\rho_{meas}$ is
  the state after the measurement, $\rho_{sys}$ is the corrected state
  of the system qubit after the QEC cycle and $\rho_{rec}$ is the
  state of the register after the full correction. (b) Circuit
  representation of the error correction algorithm. $D$ is a unitary
  operation that commutes with phase flips. $U$ is an arbitrary
  unitary operation. These operations do not affect the error
  correction functionality. }
\label{fig:schematic}
\end{figure}

A measurement in the computational basis states $|0\rangle, |1\rangle$
causes a projection onto the $\sigma_z$ axis of the Bloch sphere and
can be interpreted as an incoherent phase flip. Thus, any protocol
correcting against phase-flips is sufficient to reverse measurements
in the computational basis. The repetition code can be modified to
protect against such phase-flip errors by a simple basis change from
$|0\rangle, |1\rangle$ to $|\pm\rangle = 1/\sqrt{2}(|0\rangle \pm
|1\rangle)$. After this basis change each individual qubit is in an
equal superposition of $|0\rangle$ and $|1\rangle$ and therefore it is
impossible to gain any information about the encoded quantum
information by measuring a single qubit along $\sigma_z$. Because the
repetition code relies on a majority vote on the three-qubit register
the measurement can be only perfectly corrected for if it acts on a
single qubit as outlined in the schematic circuit shown in
Fig.~\ref{fig:schematic}(a).

This process protects the information on the system qubit, leaving it
in the same state as prior to the encoding. A complete reversal of the
measurement brings the register back to the state it had immediately
before the measurement. Therefore one needs to re-encode the register
into the protected state. This is not directly possible because the
ancilla qubits carry information about the measurement
outcome. Therefore the auxiliary qubits have to be re-initialized
prior to re-encoding as outlined in Fig.~\ref{fig:schematic}(a).

The experiment is realized in a linear chain of $^{40}$Ca$^{+}$ ions
confined in a macroscopic linear Paul trap\cite{fsk-decoherence}. Each
ion encodes a qubit in the $4S_{1/2}(m=-1/2) = |1\rangle$ and the
metastable $3D_{5/2}(m=-1/2) = |0\rangle$ state. Coherent
manipulations of the qubit state are performed by exactly timed laser
pulses in resonance with the energy difference between the two levels.
A typical experimental sequence consists of (i) initialization of the
quantum register, (ii) coherent state manipulation, and (iii)
measurement of the register. Initializing the register consists of
preparing the electronic state of the ions in a well defined state and
cooling the common motional mode of the ions close to the ground
state. In our experiment, any coherent operation can be implemented
with a universal set of gates consisting of collective spin flips,
phase shifts of individual qubits and collective entangling
operations~\cite{Volckmar,molmer}.

\begin{figure}[t]
\centering
\includegraphics[width=8.9cm]{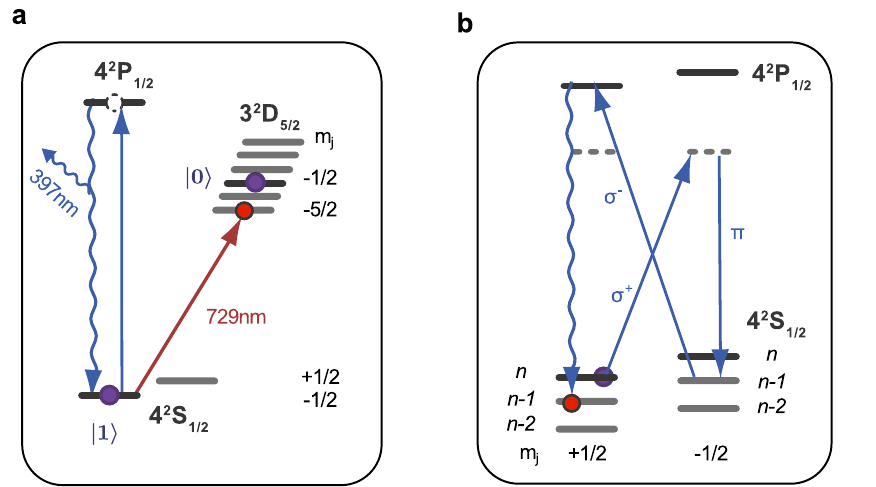}
\caption{ (a) Schematic of the measurement process on the $S_{1/2}
  \leftrightarrow P_{1/2}$ transition. The auxiliary qubits are hidden
  from the measurement by transferring the population to the $m =
  -5/2$ substate of the $D_{5/2}$ level. (b) Schematic of the Raman
  recooling procedure. This scheme utilizes two 1.5GHz detuned Raman
  beams that remove one phonon upon transition from the Zeeman
  substates $m=-1/2$ to $m=+1/2$ and an additional resonant beam
  that is used to optically pump from $m=+1/2$ to $m=-1/2$ via the
  $P_{1/2}$ state.  }
\label{fig:raman}
\end{figure}

The qubit can be measured in the computational basis by performing
electron shelving on the short-lived $S_{1/2} \leftrightarrow P_{1/2}$
transition as sketched in Fig.~\ref{fig:raman}(a).  Here, projection
onto the state $|1\rangle$ enables a cycling transition and scatters
many photons if the detection light is applied, whereas after
projection onto $|0\rangle$ no population is resonant with the laser
light at 397~nm. The outcomes can be distinguished by shining in the
laser light long enough to detect multiple photons with a
photo-multiplier tube after projecting into $|1\rangle$. The absence
of photons is then interpreted as outcome $|0\rangle$.  Although the
projection is already performed after scattering a single photon, it
is necessary to detect multiple photons for faithful discrimination.

For the reversal scheme as shown in Fig.~\ref{fig:schematic}(a) only a
single ion of the register is measured. This is realized by protecting
the other two ions from the detection light by transferring the
population from $|1\rangle$ in the $m=-5/2$ Zeeman substate of the
$D_{5/2}$ level with the procedure outlined in
Fig.~\ref{fig:raman}(a)~\cite{mark_tele}. Then, a projective
measurement does not affect the electronic state of the hidden ions
which are the remaining carriers of the information.  The uncertainty
of the measurement on the remaining ion depends on how many photons
are detected if the state was projected into $|1\rangle$.  Given that
the number of detected photons follow a Poissonian distribution, the
detection uncertainty can be easily calculated via the cumulative
distribution function of the Poisson distribution and the measurement
durations as shown in columns one to three in table~\ref{tab:results}.

The quality of subsequent coherent operations is significantly lowered
by the recoil of the scattered photons heating the motional state of
the quantum register.  Therefore, recooling the ion-string close to
the ground state is required without disturbing the quantum
information in the non-measured qubits.  In ion-traps this can be
achieved with sympathetic cooling using a second ion species.  As
trapping and cooling two different ion species requires major
experimental effort, we employ a recooling technique that can be used
with a single trapped species. We perform a Raman cooling scheme as
shown in Fig.~\ref{fig:raman}(b) while the ancilla qubits are still
protected.

Encoding and decoding of the register as shown in
Fig.~\ref{fig:schematic}(b) are implemented in our setup as described
in Ref.\cite{Schindler2011Experimental}. The encoding is realized with
a single entangling operation and the decoding is performed using a
numerically optimized decomposition into available
operations~\cite{Volckmar}. In order to facilitate the optimization
procedure, the QEC algorithm is slightly modified without affecting
its functionality by two additional unitary operations $D$ and $U$ as
shown in Fig.~\ref{fig:schematic}(b). The actual implementation can be
benchmarked with the aid of quantum state and process tomography
\cite{mark_tele,nielsen_chuang}. We use a maximum likelihood algorithm
to reconstruct the density matrix and perform a non-parametric
bootstrap for statistical error analysis~\cite{Jezek2003Quantum}.
%[Hradil,]
Because the error correction protocol acts as a single qubit quantum
channel, it can be characterized by a quantum process tomography on
the system qubit (indicated as $\rho_{sys}$ in Fig~1(a)). This
process is characterized by the process matrix $\chi_{exp}$ and its
performance compared to the ideal process $\chi_{id}$ is given by the
process fidelity $F^{proc}=\mathrm{Tr} (\chi_{id} \cdot \chi_{exp})
$. The process fidelity of a single error correction step without
measurement and recooling was measured to be $F=93(2)\%$. The process
including the measurement can be analyzed by either ignoring the
measurement outcome or by investigating the process depending on the
outcome as presented in Table~\ref{tab:results}.

% \begin{figure}[t]
% \centering
% \includegraphics[width=8.9cm]{images/chis_detection.pdf}
% \caption{(a) Histogram of measured photon counts for a measurement
%   duration of $\tau=200 \mu s$ with the corresponding $\chi$ matrices
%   for projection either in $|0\rangle$ or $|1\rangle$.  (b) Absolute
%   value of the reconstructed process matrices of the system qubit
%   after the correction step depending on the measurement outcome.}
% \label{fig:results}
% \end{figure}

The overall performance of the reversal process is determined by the
quality of the operations and the loss of coherence during the
measurement and the recooling process.  As the quality of the
operations is affected by the motional state of the ion-string after
recooling, there is a trade-off between their fidelity and the loss of
coherence during measurement and recooling. It should be noted that
the measurement affects the motion only if it is projected into the
$|1 \rangle$ state whereas the loss of coherence affects both possible
projections. The performance of the algorithm for different
measurement and recooling parameters is shown in
Table~\ref{tab:results}. A detection error of less than $0.5 \%$ is
achieved with a measurement time $\tau_{meas}=200 \mu s$ and a
recooling time of $\tau_{recool}= 800 \mu s$ leading to a mean process
fidelity of $F=84(1)\%$ which exceeds the bound for any classical
channel of $F = 50 \% $. We analyzed the measurement outcome for
$\tau_{meas}=200 \mu s$ and a measurement threshold of three photon
counts to show that no information about the encoded quantum
information can be gained by measuring a single qubit. The measurement
was performed for the initial basis states $|0\rangle, \; |0\rangle +
|1\rangle, \; |0\rangle + i |1\rangle , \; |1\rangle $ and results in
probabilities to find the outcome in state $|0\rangle$ of $48(1)\%,
50(1)\%, 50(1)\%, 50(1)\% $. This shows that indeed no information
about the initial quantum state can be inferred by measuring a single
qubit.

\begin{table*}
\begin{tabular}{c c c | c | c c c | c c c}
  $ \tau_{Raman}$ & $ \tau_{meas}$ & Detection error & $ \langle n_{phonon} \rangle$ & $F^{proc}_{mean}$ & $F^{proc}_{|1\rangle}$ & $F^{proc}_{|0\rangle}$ & $F^{rho}_{mean}$ & $F^{rho}_{|1\rangle}$ & $F^{rho}_{|0\rangle}$ \\
  \hline
  800 & 100 & 4 \%   &   0.17 &       86(3) &       82(3) &       90(2) &       89(1) &       87(1) &       91(2) \\
  800 & 200 & $<$ 0.5 \% &   0.24 &       85(2) &       87(3) &       90(3) &       84(1) &       82(1) &       85(2) \\
  800 & 300 & $<$ 0.5 \% &   0.41 &       81(3) &       78(2) &       87(3) &       84(1) &       80(1) &       87(2) \\
  800 & 400 & $<$ 0.5 \% &   0.50 &       78(3) &       71(5) &       85(4) &       82(1) &       76(2) &       90(2) \\
\end{tabular}
\caption{Columns 1 to 3: Raman recooling and measurement duration in $\mu s$ with corresponding
  detection error. Column 4: Measured mean phonon number $\langle n \rangle$ after measurement and recooling.
  Columns 5 to 7: Measured process fidelities on the system qubit without re-encoding $F^{proc}$ in(\%) and columns 8 to 10:
  Overlap of the quantum state after the full reconstruction with the state prior to the measurement $F^{rho}$ in (\%). Lower indices
  $F_{mean}$ indicate a mean fidelity while ignoring the measurement outcome. $F_{|0\rangle}$ and $F_{|1\rangle}$ 
  indicate fidelities if the measurement outcome was $|0\rangle$ ($|D\rangle$) and $|1\rangle$
  ($|S\rangle$). Errors correspond to one standard deviation.}
\label{tab:results}
\end{table*}

The presented procedure is able to protect the quantum information on
the system qubit in the presence of a quantum measurement.  In order
to perform the full measurement reversal, the ancilla qubits have to
be reset before applying the same encoding as demonstrated in
Ref~\cite{Schindler2011Experimental}.

\begin{figure*}[t]
\centering
\includegraphics[width=18.3cm]{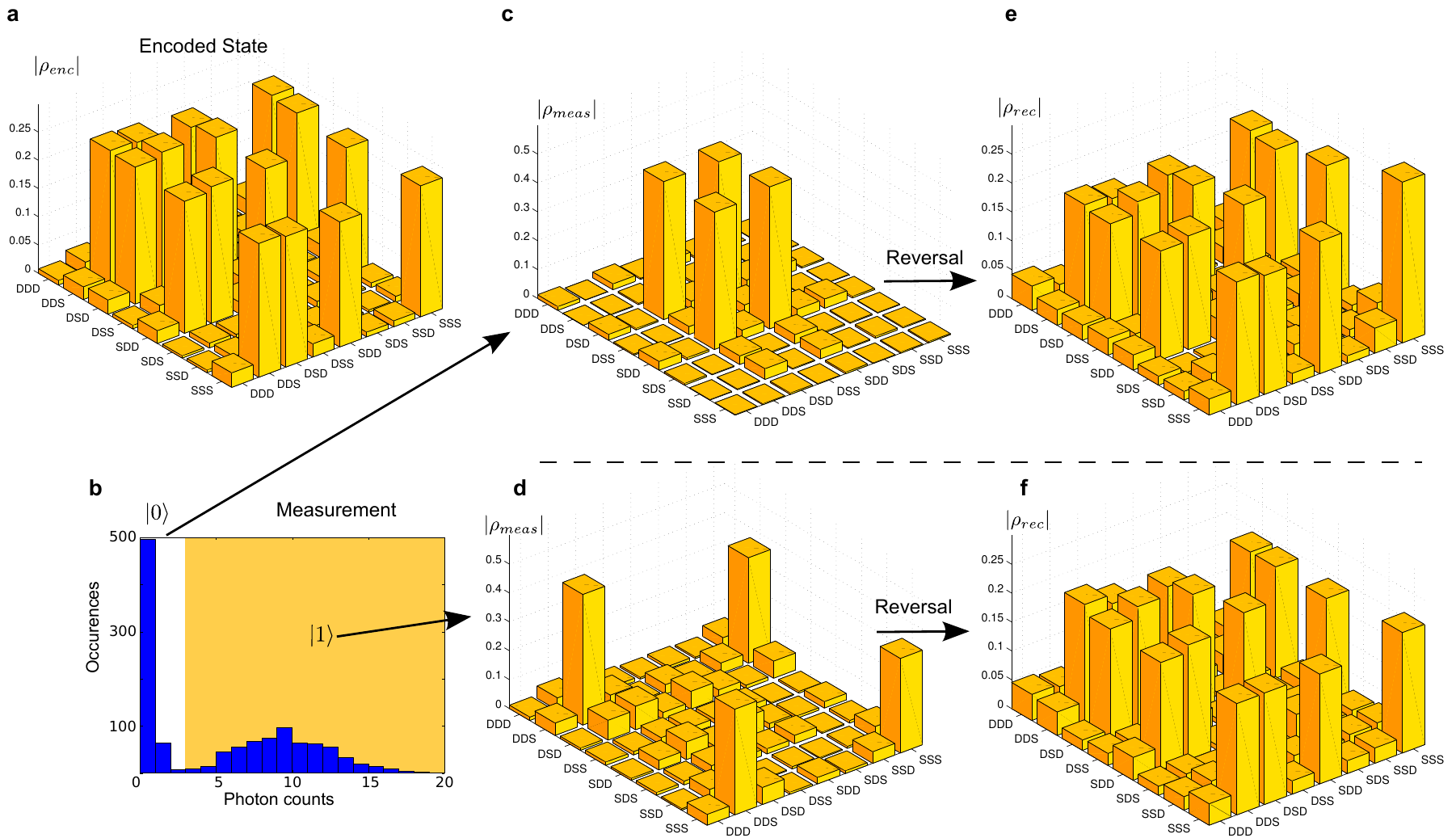}
\caption{(a) Absolute value of the reconstructed three qubit density
  matrices before the measurement $\rho_{enc}$. (b) Histogram of the
  measured photon counts for a measurement time of $200 \mu
  s$. Absolute value of three qubit density matrices after the
  measurement $\rho_{meas}$ for outcome (c) $|0\rangle$ and (d)
  $|1\rangle$. Density matrices after the measurement reversal
  $\rho_{rec}$ for outcome (e) $|0\rangle$ and (f) $|1\rangle$.}
\label{fig:results_rho}
\end{figure*}

As this technique recovers the state of the entire register, the
measurement reversal can be directly benchmarked by comparing the
state before the measurement and after the reconstruction. A quantum
state can be analyzed using quantum state tomography and evaluating
the fidelity between two states $\rho_1 , \rho_2 $ with the Uhlmann
fidelity\cite{Jozsa1994Fidelity} $F^{rho}(\rho_1, \rho_2) = (
\mathrm{Tr} \; \sqrt{\sqrt{\rho_1} \rho_2 \sqrt{\rho_1}} )^{\,
  2} \;$.  The state $\rho_{enc}$ after encoding shows a fidelity with
the ideal state of $F(\rho_{id},\rho_{enc})=94(1)\%$.  In order to
demonstrate the effect of the measurement the states $\rho_{meas}$
after measuring and recooling, and $\rho_{rec}$ after the
reconstruction are analyzed with respect to the state
$\rho_{enc}$. The measured density matrices for these states are shown
in Fig.~(\ref{fig:results_rho}).  The overlap of the state after the
measurement $\rho_{meas}$ with the state $\rho_{enc}$ is
$F(\rho_{enc},\rho_{meas})=50(2)\% $ as expected from pure dephasing
which shows that the measurement acts as dephasing when the outcome is
ignored.  In contrast, Fig.~\ref{fig:results_rho} illustrates the
evolution of the states with known outcome. The reconstructed state
$\rho_{rec}$ after correction, reset and re-encoding is measured to
have an overlap of $F(\rho_{enc},\rho_{rec})=84(1)\%$ which indicates
that the measurement was successfully reversed.  The quality of the
measurement reversal depends again on the number of scattered photons
during the measurement and the recooling time and the optimum is also
$\tau_{detect} = 200 \mu s$.  Fidelities depending on the outcome and
for various measurement durations are displayed in
table~\ref{tab:results}.

In conclusion we have demonstrated the full reversal of a strong
quantum measurement on a single qubit. We further presented an in-sequence
recooling technique that can serve as an alternative to sympathetic
two-species cooling. This may simplify the architecture for a future
large-scale ion-trap quantum information processor.
\
\begin{acknowledgements}
We thank R. Blume-Kohout for stimulating discussion.  We gratefully
acknowledge support by the Austrian Science Fund (FWF), through the
SFB FoQus (FWF Project No. F4006-N16), by the European Commission
(AQUTE), by IARPA as well as the Institut f\"{u}r Quantenoptik und
Quanteninformation GmbH.
\end{acknowledgements}
\bibliography{qec}

\end{document}